\documentstyle[osa,twocolumn,prl,aps,graphicx]{revtex}  % DON'T CHANGE
\begin{document}                % INITIALIZE - DONT CHANGE

\draft
\twocolumn[\hsize\textwidth\columnwidth\hsize\csname
@twocolumnfalse\endcsname

\title{Properties  of SAS Josephson junctions in
SO(5) theory.}
\author{B.C. den Hertog}
\address{Department of Physics, University of Waterloo, Waterloo, Ontario, Canada N2L 3G1}
\author{A.J. Berlinsky and C. Kallin}
\address{Brockhouse Institute for Materials Research and
Department of Physics and Astronomy\\
McMaster University, Hamilton, Ontario, Canada L8S 4M1}
%\date{\today}
\maketitle
\begin{abstract}                % DON'T CHANGE THIS LINE
We derive the qualitative behavior of
superconductor-antiferromagnet-superconductor (SAS) Josephson 
junctions described by Zhang's SO(5) theory. The main differences 
between these junctions and conventional SIS junctions 
arise from the non-sinusoidal current-phase relation
derived by Demler {\it et al.} for thin SAS 
junctions. Using a simple approximation 
to this non-sinusoidal function, the current voltage relation, Shapiro 
steps, thermal fluctuation effects
and the diffraction pattern in a magnetic field are obtained.

\end{abstract}
%\pacs{74.20.De,74.50.+r,74.20.-z}
\vskip2pc]

%BEGIN PAPER HERE

Recently there has been considerable attention focused on the theory of 
Zhang \cite{Zhang}, which attempts to unify the
d-wave  superconductivity in doped Mott insulators with the 
antiferromagnetism  of  undoped ``parent'' compounds.  In this
theory, the antiferromagnetic N\'{e}el vector  and the complex
superconducting order parameter form the components of a 5-dimensional
``superspin'' vector.  The SO(5) rotational symmetry of the superspin
is explicitly  broken by anisotropy which favors antiferromagnetism
for undoped or lightly-doped  compounds and superconductivity for more
heavily doped materials. The transition from  antiferromagnetism (A)
to d-wave superconductivity (S) is analogous to the  spin-flop
transition of a uniaxial Heisenberg  antiferromagnet in a parallel
magnetic field. 

  From a phenomenological perspective, the most
striking manifestations of this theory   arise in situations
where the SO(5) order parameter can be rotated from S to A  or
vice-versa by some kind of external field.  
One way of imposing such a
field is by  proximity to a material in which the superspin direction
is strongly anchored, such as in the superconductor-antiferromagnet-superconductor
(SAS) heterostructure proposed by Demler {\it et al.} \cite{Demler}.
Here we study the properties of such a heterostructure in the presence
of an external circuit and magnetic field.
Our analysis,  based upon the SAS Josephson current-phase relation obtained
within SO(5) theory\cite{Demler}, has the potential to provide
 a series of qualitative tests for the applicability of this theory
to doped Mott insulators.

\begin{center}
\begin{figure}
\includegraphics[height=6.3cm,width=6.8cm]{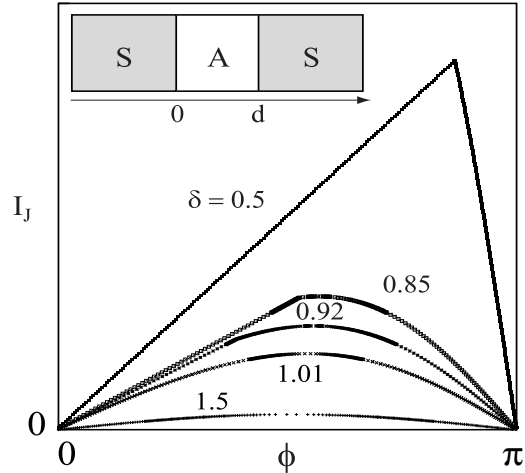}
\caption{The dc current-phase relation of an SAS junction of thickness 
$\delta=d/d_c$, as obtained in Ref.\,2.}
\label{fig1}
\end{figure}
\end{center}

The device proposed by Demler {\it et al.}   consists of a thin   
antiferromagnetic layer
sandwiched between  two strongly superconducting  regions (see inset
in  Fig.~\ref{fig1}).     If      the anisotropy    stabilizing    the
antiferromagnetism in the A region is weak, the proximity effect of the 
S regions will
align the order parameter uniformly in the S direction. 
%Demler {\it et
%al.} went on  to show   that the injection  of a supercurrent across
%the  device would reduce the proximity    effect, allowing the order
%parameter in the A  region to tip toward  the A direction.  
For the A region, there is a critical  thickness $d_c$, below which it
  remains purely  superconducting.  For thicknesses $d > d_c$,
the  order  parameter in  the A region tips  toward
 the A direction,
with   maximum tipping  angle  at  the cen-
ter of  the region,
 and  the
current-phase   relation of junctions   is similar
to that  of conventional SIS   junctions. For $d  < d_c$,  a tipping
transition  occurs at   a  critical value   of the  supercurrent or,
equivalently, of the  superconducting phase  difference  across  the
junction.  As    the phase difference is  varied   through   the critical value
$\phi_c$, the slope   of    the   current-phase    relation  changes
discontinuously,  from linear  in  the superconducting    regime for
$\phi < \phi_c$ to roughly sinusoidal for  $\phi > \phi_c$ where the 
order parameter in the A layer tips toward the A direction.

The current-phase relation  calculated in Ref. \cite{Demler} and shown
in  Fig. \ref{fig1}, can be approximated by
\begin{equation}
\label{Ij}
 I_{J}(\phi)=\left\{ \begin{array}{cr}
(\phi^{-1}_{c}I_{0}\sin\phi_{c})\phi &  0 \leq |\phi| \leq \phi_{c}\\
&\\ I_{0}\sin\phi & \phi_{c} \leq |\phi| \leq \pi \\
\end{array}
\right.
\end{equation}
where $\phi_{c}$ defines the discontinuity in the derivative of the
 current-phase relation  and $I_{0}=\phi_{c}/(\pi\delta\sin\phi_{c})$,
 while $\delta=d/d_c$ and $d_{c}/\pi=\xi_{A}$,  the antiferromagnetic
 coherence length.  It should be emphasized that although $\phi_{c}$
 is  a useful parameter for identifying a variety of physical effects
 in the SAS junction, it can  be related to more physical
 quantities. Indeed one  can  show that
\begin{equation}
\label{width}
\phi_{c}=\pi\sqrt{1 - \delta^{2}} \; . 
\end{equation}

A useful way of comparing theoretical predictions of junction behavior
to existing or future experimental data, is to analyze the behavior of
the junction  within a circuit, the classic system being the Resistively
and Capacitatively  Shunted Josephson (RCSJ) Junction\cite{Barone}. Here
 the junction is placed in  parallel with a capacitor $C$ and
resistance $R$ and a constant current is passed  through the circuit.
Such a model takes into account the capacitive effects of the
junction and also single particle tunneling  which produces a
normal state  resistance.  In the following, we explore the properties
of the SAS junction, using  the current-phase relation of
Eq. (\ref{Ij}) in an overdamped circuit where capacitive  effects are
assumed to be negligible, and we contrast these
properties to those of conventional SIS junctions which obey a
sinusoidal current phase relationship (SCPR), given by Eq. (\ref{Ij})
with $\phi_c=0$.

The dynamics of the overdamped SAS junction (ignoring thermal effects)
are governed by the equation
\begin{equation}
\label{dynamics}
\frac{\hbar}{2eR}\frac{d\phi}{dt} + I_{J}(\phi)=I_{dc} \;\,
\end{equation} 
where  $R$ is the resistance produced by single quasiparticle
tunneling  and $I_{dc}$ is an externally applied dc current.   From
Eq. (\ref{dynamics}) it is straightforward to derive the dc I-V
characteristic of the SAS junction from the Josephson equation
$(\hbar/2e)d\phi/dt=V(t),$ where $V(t)$ is the induced time dependent
voltage:
\begin{equation}
\frac{V_{dc}}{R}=\left( \frac{1}{2\pi}\int^{\pi}_{-\pi}\frac{d\phi}
{\left[I_{dc} - I_{J}(\phi)\right]}\right)^{-1}  \; .
\end{equation}

A dc  voltage is not induced in the system unless the driving current
is greater  than $I_{th}$, the  maximum of the Josephson current
$I_{J}$ \cite{classical}.  However, unlike a  junction with a SCPR,
the SAS system exhibits two distinct conditions  for this to occur,
depending on the junction  thickness. For a thin junction with
$\phi_{c}>\pi/2$, the minimum current required to induce a dc voltage is 
$I_{th}=I_{0}\sin\phi_{c}$, 
% a dc voltage is not induced unless
%$I>I_{0}\sin\phi_{c}$, 
while for a  thick junction with
$\phi_{c}<\pi/2$, a dc voltage appears only for $I_{th}=I_{0}$. 

For an SAS system with $\phi_c>\pi/2$, in the region
$I_{0}\sin\phi_{c}<I_{dc}<I_{0}$, the dc voltage is given by
\begin{eqnarray}
\label{v1}
\frac{2\pi I_{0}R}{V_{dc}}&=&\frac{1}{\sqrt{1
-\tilde{I}^{2}}}\left(\ln\left[ \frac{-\tilde{I}\tan(\phi_{c}/2) -
\sqrt{1-\tilde{I}^{2}} -1}{-\tilde{I}\tan(\phi_{c}/2) + \sqrt{1
-\tilde{I}^{2}} -1}\right]\right. \nonumber \\ &-&\left. \ln\left[
\frac{\tilde{I}\tan(\phi_{c}/2) - \sqrt{1-\tilde{I}^{2}} -
1}{\tilde{I}\tan(\phi_{c}/2) + \sqrt{1 - \tilde{I}^{2}}
-1}\right]\right) \nonumber \\ &+&
\frac{\phi_{c}}{\sin\phi_{c}}\ln\left(\frac{\tilde{I} +
\sin\phi_{c}}{\tilde{I} - \sin\phi_{c}}\right)
\end{eqnarray}
\begin{center}
\begin{figure}
\includegraphics[width=8.6cm]{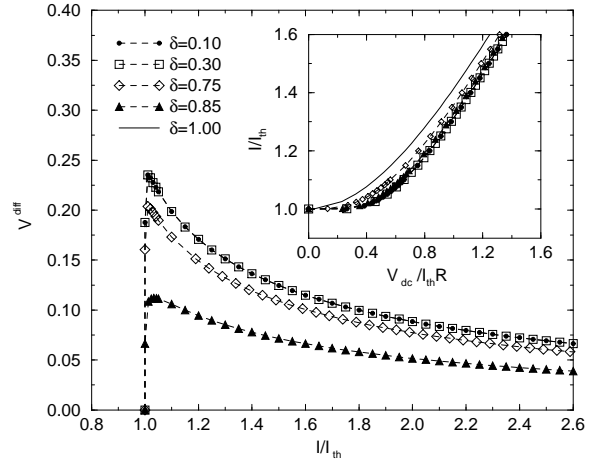}
\caption{In the  inset we show the dc I-V characteristic of an SAS junction
of   thickness $\delta$ within the  RCSJ model and assuming a
negligible capacitance.  The main body shows  for different barrier 
thicknesses the difference between  I-V curves of an SAS junction and a 
normal  junction.}
\label{ivsas}
\end{figure}
\end{center}
where  $\tilde{I}=I_{dc}/I_{0}$,  while in an SAS system with
$I_{dc}>I_{0}$,
 irrespective of $\phi_{c}$ the analogous result is
\begin{eqnarray}
\label{v2}
\frac{2\pi I_{0}R}{V_{dc}}&=&\frac{2}{\sqrt{\tilde{I}^{2} -1}}\left(
\pi + \arctan\left[\frac{-\tilde{I}\tan(\phi_{c}/2) -
1}{\sqrt{\tilde{I}^{2}-1}}\right]\right. \nonumber \\
&-&\left. \arctan\left[\frac{\tilde{I}\tan(\phi_{c}/2) -
1}{\sqrt{\tilde{I}^{2}-1}}\right]\right) \nonumber \\
&+&\frac{\phi_{c}}{\sin\phi_{c}}\ln\left(\frac{\tilde{I} +
\sin\phi_{c}}{\tilde{I}-\sin\phi_{c}}\right) \; .
\end{eqnarray}

There are several comments to be made on the two expressions
above. For a junction with a SCPR, the I-V relation has the form
$V^{n}_{dc}=RI_{0}\sqrt{\tilde{I}^{2}-1}$ \cite{Aslamazov},  which is
finite for $I_{dc}>I_{0}$, and shows ohmic behavior as the driving
current $I_{dc}$ becomes large.  For $0\le\phi_{c}<<\pi/2$,
Eq. (\ref{v2}) reduces to this  form, since in this limit the SAS
current-phase relation is  well approximated by the normal Josephson
expression. On the other hand, there  is a  clear  qualitative
difference between Eqs. (\ref{v1}) and (\ref{v2}) as $\phi_{c}$
becomes larger. For $\phi_{c}>\pi/2$, as $I_{dc}\rightarrow
I_{0}\sin\phi_{c}$, $V_{dc}$ tends to zero logarithmically. This
implies that as the driving current is increased from zero, there
should be a very rapid increase in the induced dc voltage as soon as
$I_{dc}$ rises above $I_{0}\sin\phi_{c}$.

By plotting  $V_{dc}/I_{th}R$ against $I/I_{th}$, and  relating
$\phi_c$ to $\delta$ through  Eq. (\ref{width}), we can compare I-V
curves  for SAS systems of  various barrier widths. This comparison is
shown in the inset of Fig. \ref{ivsas}.  Note that as
$\delta\rightarrow 1\;(\phi_{c}\rightarrow 0)$, the I-V curve  becomes
identical to that of a regular junction with a SCPR. On the other
hand, as the barrier width becomes narrower, the dc voltage develops a
sharp  (logarithmic) leading edge and clearly deviates from
nor-
\begin{center}
\begin{figure}
\includegraphics[width=7.0cm]{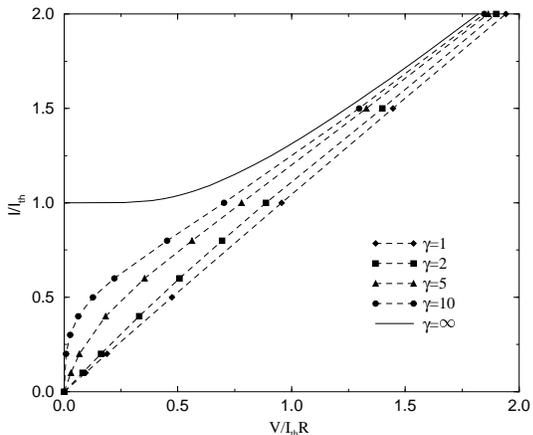}
\caption{ Thermal I-V characteristics  for an SAS junction of
thickness $\delta=0.1$ for a range of $\gamma=I_{th}\hbar/ekT$.}
\label{thermal}
\end{figure}
\end{center}
mal
junction behavior. These differences will show  up most clearly in the
differential resistivity.  In the body of Fig. \ref{ivsas} we plot the 
difference between  I-V curves   of the SAS junction  and the normal
junction,  defined as $V^{\rm{diff}}=V_{dc}/I_{th}R -V_{dc}^{n}/I^{n}_{th}R$. 
 The maximum deviation  from normal junction
behavior occurs  when the driving current $I_{dc}$ is just above the
threshold current $I_{th}$.

Thermal effects lower the threshold current to zero, interpolating
between the $T=0$ behavior and a linear I-V relation as was shown by
Ambegaokar and Halperin \cite{ah}. The analogous results are shown in
Fig. \ref{thermal} for a thin SAS junction \cite{sheehy}.  Comparison to
Ref. \cite{ah} shows that the largest differences between SAS and
conventional junctions occur at low temperatures and small voltages.

The addition of an ac component to the dc driving current produces
Shapiro  steps\cite{Shapiro} in the I-V characteristics of RCSJ
systems at certain  values of the dc voltage \cite{Waldram}. These
steps of constant dc voltage  for a range of values of the dc driving
current can be understood as arising  from  phase-locking between the
fundamental Josephson frequency of the junction  or one of its
harmonics and a harmonic of the applied ac current. This leads us to
consider the  addition of an ac  driving current to the SAS system
described above, since  the unusual form of the   current-phase
relation may lead to features within the step structure of  the dc
I-V characteristic. In particular, the time dependent current
$I_{J}(\phi[t])$ induced  in the SAS junction from the purely dc
driving current, should have a rich harmonic structure which can
respond to the  harmonics  of the additional ac current. 

In order to explore this point,  we add to the right hand side of
Eq. (\ref{dynamics})  an applied  current of the form
$I_{ac}\sin\omega t$.  By making a change of variables $u=\omega t$
and defining $A_{x}=2eI_{x}R/\hbar\omega$  where the subscript $x=\{dc,ac,J,th\}$,
the dynamics of our circuit can be described by 
\begin{equation}
\label{shapiro}
\frac{d\phi}{du} + A_{J} = A_{dc} + A_{ac}\sin u \; ,
\end{equation}
which can be solved numerically to obtain $\langle d\phi/du\rangle
=2eV_{dc}/\hbar\omega$ as a function of $A_{dc}$, for fixed values of
$A_{th}$ and $A_{ac}$. Steps are expected to occur when $\langle
d\phi/du\rangle=n$  ($V_{dc}=n \hbar  \omega/2e$) \cite{Waldram},
where  $n$ is a non-negative integer. This corresponds to the
previously mentioned  phase-locking between the fundamental Josephson
frequency and a harmonic of the  applied ac current. In addition, sub-steps should appear when  $\langle d\phi/du\rangle=n/m$, due to
phase-locking between harmonics of the Josephson  frequency and the
applied ac current frequency.  

\begin{center}
\begin{figure}
\includegraphics[width=8.6cm]{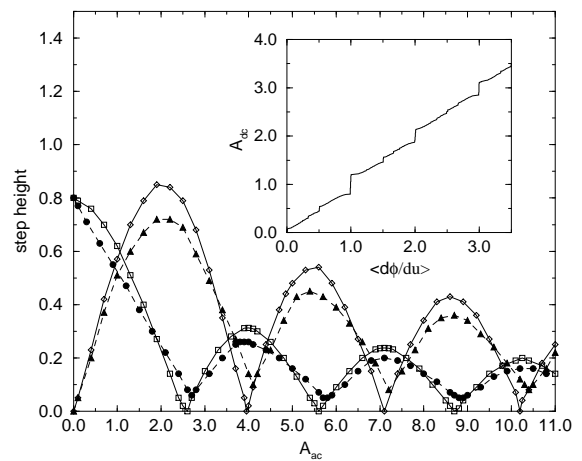}
\caption{In the inset  we show the I-V characteristic of an SAS junction
with barrier thickness  $\delta=0.5, A_{th}=0.8$  and $A_{ac}=6.0$. 
 The main body  compares the variation in zero step-heights  produced by an 
SAS junction (circles) to a normal junction with  a SCPR (squares) for 
different values of $A_{ac}$. Also compared are the first step-heights of  
SAS (triangles) and   normal (diamonds) junctions. }
\label{shapirofig}
\end{figure}
\end{center}

From our numerical analysis of
Eq. (\ref{shapiro}), we show in the inset of  Fig. \ref{shapirofig}
a  typical  stepped I-V characteristic for a narrow  ($\delta=0.5$)
SAS junction with parameters $A_{th}=0.8$ and $A_{ac}=6.0$.
The body of Fig. \ref{shapirofig} shows the change in  step-heights as the ac
driving current amplitude is varied (dashed lines). The zeroth  step-height is
defined as the value of $A_{dc}$ at which $\langle d\phi/du\rangle$
becomes non-zero, while the first step-height is simply the length of
the vertical step at $\langle d\phi/du\rangle=1$, and so on (see inset).  
When $A_{ac}=0$ , the  zeroth step-height
is simply given by $A_{th}=2eRI_{th}/\hbar \omega$, where $I_{th}$ is
the threshold current required to induce a dc voltage  as defined
earlier. In this figure we also compare the zeroth  and first
step-heights of the SAS junction to   a normal
junction  with a SCPR and  the same value of $A_{th}$ (solid lines).

There are several distinct features  which distinguish the SAS Shapiro
step  behavior from that of a regular Josephson system. As can be seen
in the figure,  there is a clear linear behavior in the zeroth
step-height  for small values of $A_{ac}$ for the  SAS junction.  For
a junction with a SCPR, the  step-height dependence on $A_{ac}$  has a
quasi-Bessel function behavior where the Bessel function is of the same
order  as the step-height \cite{Waldram,Russer}. 
For the SAS junction,
it would appear that the higher harmonic con-
\vspace{-5mm}
tent  of $I_{J}$
(Eq.\,\ref{Ij}), allows phase-locking to occur for
 all  values of
$A_{ac}$, in contrast to a regular junction with a SCPR, where there
are zeroes in the step-heights for certain values of
$A_{ac}$. 
Furthermore, the maxima of the step-heights are suppressed
for the $SO(5)$ case, and there appears to be a slight shift in the
periodicity of the  step-height behavior. 
It is worth noting  too,
that from the  higher harmonic content of the SAS current phase
relation, there are large step-heights  (sub-steps) at non-integer values of
$\langle d\phi/du\rangle$. These are enhanced over what  would be
expected in the  Shapiro steps produced by a regular junction with a
SCPR.

As a last illustration of the properties of SAS junctions, we
consider the  isolated junction in the presence  of a magnetic field.
A magnetic field will produce Fraunhofer diffraction peaks for  the
maximum current through an SAS junction as a function of the magnetic
flux threading  through it. For the SAS junction in a magnetic field,
we consider the system to be  lying flat along the $z$-axis with
thickness $d$, a height $h$ along the $y$-axis and a  depth $L$
parallel to the $x$-axis. Applying a magnetic field $H$ along the
positive  $y$ direction implies that the gauge-invariant
superconducting phase  difference across the junction is given by
$\phi[x]=2\pi Hz_{0}x/\Phi_{0} + \phi_{0}$, where $\phi_{0}$ is the
phase difference at one end of the  junction $(x=0)$ and $\Phi_{0}$ is
the flux quantum \cite{Tinkham}. In general, the current, $I$  through the 
junction will be given by $\int_{0}^{L}J_{J}(\phi[x]) dx$ where the  integrand
is the current density  $I_{J}(\phi[x])/L$.  For a junction with
$\phi_{c}>\pi/2$, one can show that the current is given by
\begin{equation}
\label{imax}
I=\frac{\Phi_{0}I_{0}}{2\pi\Phi}\left[\frac{\sin\phi_{c}(\phi_{c}-\phi_{0})^2}{2\phi_{c}}-
\cos\left(\phi_{0} + {2\pi\Phi\over\Phi_{0}}\right)
+\cos\phi_{c}\right] 
\end{equation}
where $\Phi=Hz_{0}L$ is the magnetic flux threading the junction. The
maximal current  $I_{max}$ is obtained by maximizing $I$ with respect
to $\phi_{0}$. This leads to the  following equation for $\phi_{0}$,
\begin{equation}
\label{phi0}
\sin(\phi_{0} +
2\pi\Phi/\Phi_{0})=\frac{\phi_{0}\sin\phi_{c}}{\phi_{c}} .
\end{equation}

On the other hand, for a junction with $\phi_{c}<\pi/2$, $I_{max}$ is
given  by the regular Fraunhofer expression
$I_{max}=I_{0}|\sin(\pi\Phi/\Phi_{0})|/ (\pi\Phi/\Phi_{0})$ for
$\phi_{c}<|\phi_{0}|<\pi/2$, and then is determined by
Eqs. (\ref{imax}) and (\ref{phi0}) for $|\phi_{0}|<\phi_{c}$.

\begin{center}
\begin{figure}
\includegraphics[width=7.6cm]{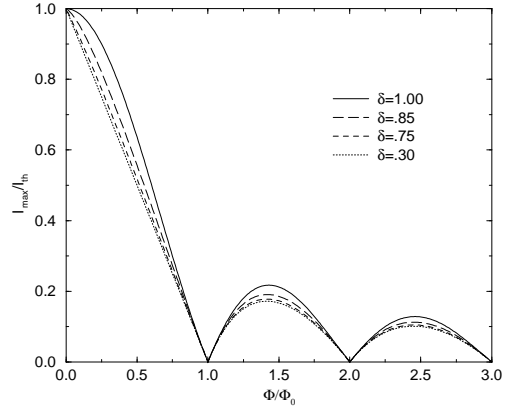}
\caption{Diffraction pattern of an SAS junction  with  barrier thickness 
$\delta$. The regular Fraunhofer pattern from a normal junction
 is given by the $\delta=1$ curve.}
\label{magfield}
\end{figure}
\end{center}

We have determined the maximal current through the SAS junction as a
function  of magnetic flux $\Phi/\Phi_{0}$ by systematically solving
Eqs. (\ref{imax})  and (\ref{phi0}) for increasing  values of the
magnetic  flux ratio $\Phi/\Phi_{0}$.  In Fig.~\ref{magfield} we  show
the variation of $I_{max}/I_{th}$ with the magnetic flux ratio  for a
variety of SAS barrier  thicknesses.  Fig.~\ref{magfield} shows a
general suppression of the diffraction peaks and a  strong linear
dependence before the first diffraction minimum, reminiscent of the Shapiro step-height  behavior.

In summary, we propose that the existence of an SO(5) superspin vector
can be inferred from distinctive features
\vspace{-2mm}
 of an overdamped  SAS junction,
particularly  the rapid
 increase of
 the voltage in the low T I-V curve, the linear 
dependence of Shapiro step heights as a function of ac driving current, and
the linear dependence of the critical current at low magnetic fields.

The authors would like to thank E. Demler for useful discussions.
This  work was supported in part by the
 Natural Sciences and
Engineering Research Council of Canada, by the Ontario Center for
Materials Research and by Materials and Manufacturing Ontario. 
BCdH would like to thank the  BIMR, McMaster University for their hospitality 
during the period of this work.

% \begin{figure}  % Please send figures with disk, or separately if
% if it is an e-mail submission. (Good photo or India ink drawing.)
% \caption{Please place your figure caption here.}
% \end{figure}

% \begin{table}
% \caption{Please place your table caption here.}
% \begin{tabular}{lrcd} % In second brace, l = left, r = right,
% c = centered and d = decimal justification.
% One&Two&Three&Four\\  % Separate items with &. End line with \\
% \tableline % Creates a horizontal line.
% One&Two\tablenote{footnote.}&Three&Four\\ % Place \tablenote{}
% after item to be footnoted.
% \end{tabular}
% \end{table}


\begin{references}
\bibitem{classical} This is simply the analogy of a massless particle
moving along the $\phi$ axis in a potential defined by $\int (I_{dc} -
I_{J}) d\phi$. For $I<I_{th}$, there will be a local minimum
where the particle will come to rest.
\end{references}
\end{document}